\documentclass[10pt]{article}

\usepackage{cite,makeidx,amscd,latexsym,graphicx,graphics,shapepar,hyperref,amsmath,amsthm,amsopn,amstext,amsfonts,amsbsy}

\makeindex \textwidth=470pt 
\input{epsf}

\begin{document}

\author{ Alireza Khalili Golmankhaneh $^{a\dag}$\\Dumitru Baleanu $^{b,c,d}$
\footnote{Tel:+903122844500, ~~~~Fax:+903122868962~\newline
\textit{E-mail addresses}:~dumitru@cankaya.edu.tr~}
\\$^a$\textit{Department of Physics,~Islamic Azad University, Urmia
Branch,}\\\textit{PO Box 969,~Urmia,~Iran}\\
$^\dag$E-mail:alirezakhalili2005@gmail.com\\
$^{b}$\textit{Department of Mathematics and Computer Science}\\
\textit{$\c{C}ankaya$} \textit{University, 06530 Ankara,
Turkey}\\ $^{c}$\textit{Institute of Space Sciences,}\\
\textit{P.O.BOX, MG-23, R76900,~Magurele-Bucharest, Romania}  \\
$^{d}$\textit{Department of Chemical and Materials Engineering, Faculty of
Engineering}, \\ \textit{King Abdulaziz University, P.O. Box: 80204, Jeddah, 21589,
Saudi Arabia}}

\title{\textbf{ Schr\"{o}dinger Equation on Fractals Curves Imbedding in $R^3$}}\maketitle \large

 \begin{abstract}
 In this paper we have generalized the quantum mechanics on fractal time-space.~The time is changing  on Cantor-set like but space is considered as fractal curve like Von-Koch curve.~The Feynman path method in quantum mechanics has been suggested on fractal curve.~ Using $F^{\alpha}$-calculus and Feynman path method we found the Schr\"{e}dinger on fractal time-space.~The Hamiltonian operator and momentum operator has been derived. More, the continuity equation and the probability density is given in generalized formulation.
   \end{abstract}
 \\
\section{Introduction}
 Fractal is objects that are very fragmented and irregular at all scales.~Their important properties are non-differentiability and having non-integer dimension.~Fractal has topological dimension   less than Hausdorff-Besicovitch, box-counting, and similarity dimensions.~In general, dimension of fractal can be integer or not well-defined dimension\cite{book-1,book-2,book-3,book-4,book-5,book-6,book-7}. Fractional local calculus and nonlocal has applied to model
  the process with memory and fractal structure\cite{Fractional-1,Fractional-2,Fractional-3,Fractional-4,Fractional-5,Fractional-6,Fractional-7}.
  The electric and magnetic fields are derived using fractional integrals as a approximation method on fractals \cite{tra-1}. The quantum space-time on the basis of  relativity principle and geometrical concept of fractals is introduced  \cite{nottale-1}.The probability density of quantum wave function with by Dirichlet boundary conditions in a D-dimensional spaces has been studied \cite{Berry-1}. The fractal concept to quantum physics and the relationships between fractional integral and  Feynman path integral method is developed \cite{Laskin-1,Laskin-2}. The generalized wave functions is introduced to fractal dimension, a wide class of quantum problems, including the infinite potential well, harmonic oscillator, linear potential, and free particle \cite{Wojcik-1}.Fratal path in quantum mechanics and their contributing in Feynman path integral is investigated \cite{Amir-1}. The classical mechanics is derived without the need of the least-action principle using path-integral approach \cite{Gozzi-1}.
  The calculus on the fractals has been studied in different methods like probabilistic approach method, sequence of discrete Laplacians,  measure-theoretical method,  time scale calculus  \cite{Kigami-1}.Riemann integration like method has been studied since that is  useful and algorithmic \cite{fractal-Gangal1,fractal-Gangal2,fractal-Gangal3,fractal-Gangal4,fractal-Gangal5,fractal-Gangal6}.Using the calculus on fractals the Newtonian mechanics,  Lagrange and Hamilton mechanics, and Maxwell equations has been generalized \cite{fractal-Gangal7,fractal-Gangal8,fractal-Gangal9}. As a pursue theses research
  we generalized the quantum mechanics on fractals.\\
  The plan of this paper is as following:\\
    Section \ref{1-s} we  review the fractal calculus. In section \ref{2-s} we defined the gradient, divergent and Laplacian on fractal space.~Section \ref{3-s} is explained the quantum mechanics
    on fractals curves.~In section \ref{4-s} we suggested the probability density and continuity equation
    on the generalized quantum formalism.~Finally, section  \ref{5-s} is devoted to conclusion.
\section{A Summery of the calculus on fractal curves\label{1-s}}
We review the $F^{\alpha}$-calculus on fractal curves \cite{fractal-Gangal1,fractal-Gangal2,fractal-Gangal3,fractal-Gangal4,fractal-Gangal5,fractal-Gangal6,fractal-Gangal7,fractal-Gangal8,fractal-Gangal9}. Suppose fractal curve $F \subset R^{3}$ which is continuously parameterizable i.e there
exists a function  $\textbf{w}:[a_{0}, b_{0}] \rightarrow F\subset R^{3}$ which is continuous. We also assume $\textbf{w}$ to be invertible. A subdivision $P_{[a,b]}$ of interval $[a,b], a<b,$ is a finite set of points $\{a= v_{0}<v_{1},...<v_{n}=b\}$. For $a_{0}\leq a<b<b_{0}$ and appropriate $\alpha$ to be chosen, therefore let
 \begin{equation}\label{fractal-1-eq}
    \gamma^{\alpha}(F,a,b)=\lim_{\delta\rightarrow 0}\inf_{\{P_{[a,b]}:|P|\leq
    \delta\}} \sum_{i=0}^{n-1}\frac{|\textbf{w}
    (v_{i+1})-\textbf{w}(v_{i})|^{\alpha}}{\Gamma(\alpha+1)},
\end{equation}
where $|.|$ denotes the Euclidean norm on $R^{3}$ and $|P|=\max \{v_{i+1}-v_{i};i=0,...,n-1\}$. A $\gamma$-dimension of $F$, which is defined as
\begin{equation}\label{fractal-2-eq}
\textmd{dim}_{\gamma}(F)=\inf\{\alpha:\gamma^{\alpha}(F,a,b)=0\}=\sup
\{\alpha:~\gamma^{\alpha}(F,a,b)=\infty\}.
\end{equation}
After this defintion $\alpha$ is equal to $dim_{\gamma} (F).$ The staircase function $S_{F}^{\alpha}: [a_{0},b_{0}]\rightarrow R$ of order $\alpha$ for a set $F$, is defined as
\begin{equation}\label{fractal-3-eq}
    S_{F}^{\alpha}(v)=\begin{cases}
\gamma^{\alpha}(F,p_{0},v)~~~v\geq p_{0}\\
-\gamma^{\alpha}(F,v,p_{0})~~~v< p_{0},\end{cases}
\end{equation}
where $a_{0}\leq p_{0}\leq b_{0}$ is arbitrary but fixed, and $v\in [a_{0},b_{0}].$ It is monotonic function.~ The $\theta=\textbf{w}(v),$ denote  a point on fractal curve $ F$
\begin{equation}\label{fractal-4-eq}
    J(\theta)=S_{F}^{\alpha}(\textbf{w}^{-1}(\theta)),~~~\theta \in F.
\end{equation}
We suppose that fractal curves whose $S_{F}^{\alpha}$ is finite and invertible on $[a,b]$. The $F^{\alpha}$-derivative of the bounded function $f: F\rightarrow R$ ~$(f\in B(F))$ at $\theta \in F$ is defined.\\\\
Then the
\textbf{directional $F^{\alpha}$-derivative} of function $f$ at $\theta\in F$ is
defined as
\begin{equation}\label{fractal-5-eq}
    ^{w_{j}}\mathfrak{D}_{F}^{\alpha}f(\textbf{w}(v))=F-\lim_{t'\rightarrow t}\frac{f(w_1(v), w_2(v),...w_j(v'),...w_i(v))-f(\textbf{w}(v))}{S_{F}^{\alpha}(v')-S_{F}^{\alpha}(v)},
\end{equation}
where $w_{j}$ is shows direction of $F^{\alpha}$-derivative, if the limit exists \cite{fractal-Gangal4}.\\
Let $f\in B(F)$ is an $F$-continuous function on $C(a,b)$ which is the segment $\{ \textbf{w}(v) : v\in [a,b]\}$ of $F$. Now let
$g: f\rightarrow R$ be define as
\begin{equation}\label{fractal-6-eq}
    g(w(v))=\int_{C(a,v)}f(\theta)d_{F}^{\alpha}\theta,
\end{equation}
for all $v\in[a,b]$. So that
\begin{equation}\label{sertqw}
    \mathfrak{D}_{F}^{\alpha}g(\theta)=f(\theta)
\end{equation}
Note:~Let $\gamma^{\alpha}(F,a,b)$ be finite and $f(\theta)=1$,
$\theta\in F$ denote the constant function. Then
\begin{equation}\label{aaa}
    \int_{C(a,b)}f(\theta)d_{F}^{\alpha}\theta=
    \int_{C(a,b)}1d_{F}^{\alpha}\theta=S_{F}^{\alpha}(b)-S_{F}^{\alpha}(a)=J((w(b))-J((w(a)).
\end{equation}
Remark: $F^{\alpha}$-derivative and $F^{\alpha}$-integral is a
linear operation.\\
1)~Let $f:F\rightarrow R$, $f(\theta)=k\in R$ then
$\mathfrak{D}_{F}^{\alpha}f=0$. \\
2)~IF $f:F\rightarrow R$ be a $F$-continuous function such that
$\mathfrak{D}_{F}^{\alpha}f=0$. Then $f=k$ where $k$ on $C(a,b).$\\
Suppose $f:F\rightarrow R$ be $F^{\alpha}$-differentiable function and
$h:F\rightarrow R$ be $F$-continuous such that
$h(\theta)=\mathfrak{D}_{F}^{\alpha}f(\theta)$, so
\begin{equation}\label{4po}
    \int_{C(a,b)}h(\theta)d_{F}^{\alpha}\theta=f(w(b))-f(w(a)).
\end{equation}
Analogue Taylor series is defined for $h(\theta)\in B(F)$ as
\begin{equation}\label{mk}
    f(\textbf{w}(v))=\sum_{n=0}^{\infty}\frac{(S_{F}^{\alpha}(v)-S_{F}^{\alpha}(v'))^{n}}
    {n!}(\mathfrak{D}_{F}^{\alpha})^{n}f(\textbf{w}(v')),
\end{equation}
where $h(\theta)$ is $F^{\alpha}$-differentiable any number of times on
$C(a,b)$. That is $(\mathfrak{D}_{F}^{\alpha})^{n}h\in B(F)$, $ \forall n>0$.
\section{Gradient,  Divergent, Curl  and Laplacian on Fractal Curves\label{2-s}}
In this section we generalized the $F^{\alpha}$-calculus by defining the gradient,  divergent, curl and Laplacian on fractal curves imbedding in $R^3$.
\subsection{Gradient on fractal curves}
Let us consider the $f\in B(F)$ as an $F$-continuous function on $C(a,b)\subset F$ and $\textbf{w}(v,w_{i}(v)): R\rightarrow R^3, i=1,2,3$, ~so the gradient of the $f(\textbf{w}):F\rightarrow R$ is
\begin{equation}\label{grad}
    \mathfrak{\nabla}_{F}^{\alpha}f(\textbf{w})= ~^{w_{i}}\mathfrak{D}_{F}^{\alpha}f(\textbf{w})\hat{e}^{i}~~~i=1,2,3,
\end{equation}
where the $\hat{e}^{i}$ is the basis of $R^n$.
\subsection{Divergent on fractal curves}
Let the $\textbf{f}(\textbf{w}(v))=f_{i}(\textbf{w}(v))~\hat{e}^{i}~~i=1,2,3$, be a vector field on fractal curve. Then we define the divergent of the $\textbf{f}: F\rightarrow R^n$ as follows:
\begin{equation}\label{ewswe}
    \mathfrak{\nabla}_{F}^{\alpha}.\textbf{f}(\textbf{w}(v))=~^{w_{i}}\mathfrak{D}_{F}^{\alpha}f_{i}(\textbf{w}(v)),
\end{equation}
where $f_{i}(\textbf{w}(v))$ are components of vector field.
\subsection{Laplacian on fractal curves}
Consider the $\textbf{w}(v,w_{i}(v)): R\rightarrow R^3$ on the fractal curve, therefore the Laplacian is defined as
\begin{equation}\label{sedr}
   \triangle_{F}^{\alpha}f=(\mathfrak{\nabla}_{F}^{\alpha})^2f= (^{w_{i}}\mathfrak{D}_{F}^{\alpha})^2f(\textbf{w}(v))
\end{equation}
where the $\triangle_{F}^{\alpha}$ is called Laplacian on fractal curve.
\section{Quantum mechanics on fractal curve \label{3-s}}
The classical mechanics is reformulated in terms of
a minimum principle. The Euler-Lagrange equations of motion is derived from the least action.~The Feynman paths  for a particle in quantum mechanics are fractals with dimension 2 \cite{Abbott1}. In this section, we obtain the Schr\"{o}dinger equation on fractal curves.
\subsection{Generalized Feynman path integral method }
 Feynman method for studying quantum mechanics using classical Lagrangian  and action is presented in Ref \cite{fey1,fey2}. Now we want to generalized Feynman  method using Lagrangian on fractals curves. Consider generalized action as
 \begin{equation}\label{semnhd}
\mathfrak{A}^{\alpha}_{F} =\int_{t_{1}}^{t_{2}}L_{F}^{\alpha}(t,\textbf{w}(v),~^tD_{F}^{\alpha}\textbf{w}(v))d_{F}^{\alpha}v~d_{F}^{\alpha}t~~~~L_{F}^{\alpha}:F\times F \times F\rightarrow R.
 \end{equation}
 In view of Feynman method, if wave function on fractal in $t_{1},\textbf{w}_{a}(v_{1})$ is  $\psi_{F}^{\alpha}(t_{1},\textbf{w}_{a}(v_{1}))$. So it gives the total probability amplitude  at $t_2,\textbf{ w}_b(v_2) $ as
 \begin{equation}\label{g}
   \psi_{F}^{\alpha}(t_{2},\textbf{w}_{b}(v_{2}))=\int_{-\infty}^{\infty}K_{F}^{\alpha}(t_{2},\textbf{w}_{b}(v_{2}),t_{1},\textbf{w}_{a}(v_{1}))
   (\psi_{F}^{\alpha}(t_{1},\textbf{w}_{a}(v_{1}))d_{F}^{\alpha}\textbf{w}(v),
 \end{equation}
 where $K_{F}^{\alpha}$ is the propagator which is defined as follows:
 \begin{equation}\label{aswq}
    K_{F}^{\alpha}(t_{2},\textbf{w}_{b}(v_{2}),t_{1},\textbf{w}_{a}(v_{1}))=\int_{w_{a}}^{w{b}}\exp[\frac{i}{\hbar}\mathfrak{A}^{\alpha}_{F}]
    \mathcal{D}_{F}^{\alpha} \textbf{w}(v).
 \end{equation}
 Where symbol $\mathcal{D}_{F}^{\alpha}$ indicates the integration over all fractal paths from $\textbf{w}_a(v_1)$ to $\textbf{w}_b(v_2)$.\\
Now we derive the Schr\"{o}dinger equation for a free particle on fractal curve, which is describes the evolution of the wave function from $\textbf{w}_{a}(v_1)$  to $\textbf{w}_{b}(v_2)$ , when $t_2$ differs
from $t_{1}$ an infinitesimal amount $\epsilon$. Supposing  $S_{F}^{\alpha}(v_2)=S_{F}^{\alpha}(v_1)+\epsilon$, leads to Lagrangian for free particle as
\begin{equation}\label{dewww}
    L_{F}^{\alpha}(t,\textbf{w}(v),~^tD_{F}^{\alpha}\textbf{w}(v))\simeq\frac{m (\textbf{w}(v)-\textbf{w}(v_0))^2}{2(S_{F}^{\alpha}(v_2)-S_{F}^{\alpha}(v_1))}.
\end{equation}
The generalized action on fractal  $ \mathfrak{A}^{\alpha}_{F}$ is approximately
\begin{equation}\label{sweaqw}
    \mathfrak{A}^{\alpha}_{F}\sim \epsilon  L_{F}^{\alpha}=  \frac{m (\textbf{w}(v)-\textbf{w}(v_0))^2}{2\epsilon}.
\end{equation}
As a consequence, we obtain
\begin{equation}\label{dddd}
    \psi_{F}^{\alpha}(t+\epsilon,\textbf{w}(v))=\int_{-\infty}^{+\infty}\frac{1}{A}\exp[\frac{i}{\hbar}\frac{m (\textbf{w}(v)-\textbf{w}_0(v_0))^2}{2\epsilon}]
    \psi_{F}^{\alpha}(t,\textbf{w}_0(v_0))\mathcal{D}_{F}^{\alpha}\textbf{w}_0(v_0).
\end{equation}
Here, because of properties of exponential function only  those fractal paths give significant contributions which are very close to $\textbf{w}(v)$. Changing the variable in the integral $\delta=\textbf{w}(v)-\textbf{w}_0(v_0)$ we have $\psi_{F}^{\alpha}(t,\textbf{w}_0(v_0))=\psi_{F}^{\alpha}(t,\textbf{w}(v)+\delta)$.~Since both $\epsilon$ and $\delta$ are small quantities, so that $\psi_{F}^{\alpha}(t,\textbf{w}(v)+\delta)$ and $\psi_{F}^{\alpha}(t+\epsilon,\textbf{w}(v))$   can be expanded using
Eq. (\ref{mk}). We only keep  up to terms of second order of the $\epsilon$ and $\delta$. As a result we get
\begin{align}\label{rewq}
    \psi_{F}^{\alpha}(t,\textbf{w}(v))+\epsilon (~^t\mathfrak{D}_{F}^{\alpha})\psi_{F}^{\alpha}(t,\textbf{w}(v))\simeq\chi_{F}(t)
    \int_{-\infty}^{+\infty}\frac{1}{A}\exp[\frac{i}{\hbar}\frac{m \delta^2}{2\epsilon}](
    \psi_{F}^{\alpha}(t,\textbf{w}(v))+ \delta (~^{w_{i}}\mathfrak{D}_{F}^{\alpha})\psi_{F}^{\alpha}(t,\textbf{w}(v)))\nonumber\\+\frac{\delta^2}{2}
    (~^{w_{i}}\mathfrak{D}_{F}^{\alpha})^2\psi_{F}^{\alpha}(t,\textbf{w}(v))) d_{F}^{\alpha} \delta,
\end{align}
where $\chi_{F}(t)$ is the characteristic function for Cantor like sets.~The second term in the right hand side vanishes on integration. It follows by equating the leading terms
on both sides we obtain
\begin{equation}\label{fffdr}
    \psi_{F}^{\alpha}(t,\textbf{w}(v))=\int_{-\infty}^{+\infty}\frac{1}{A}\exp[\frac{i}{\hbar}\frac{m \delta^2}{2\epsilon}]\psi_{F}^{\alpha}(t,\textbf{w}(v))d_{F}^{\alpha} \delta.
\end{equation}
Also, we arrive at
\begin{equation}\label{dewsf}
    A=\int_{-\infty}^{+\infty}\frac{1}{A}\exp[\frac{i}{\hbar}\frac{m \delta^2}{2\epsilon}]d_{F}^{\alpha} \delta=\sqrt{\frac{2 i \pi \hbar \epsilon}{m}},
\end{equation}
and
\begin{equation}\label{ewqwe}
    \int_{-\infty}^{+\infty}\frac{1}{A}\exp[\frac{i}{\hbar}\frac{m \delta^2}{2\epsilon}](
    \frac{\delta^2}{2}
    (~^{w_{i}}\mathfrak{D}_{F}^{\alpha})^2\psi_{F}^{\alpha}(t,\textbf{w}(v))) d_{F}^{\alpha} \delta=\epsilon\frac{i \hbar}{2 m}(~^{w_{i}}\mathfrak{D}_{F}^{\alpha})^2\psi_{F}^{\alpha}(t,\textbf{w}(v))).
\end{equation}
Finally, equating the remaining terms, we get Schr\"{o}dinger equation on fractal curves for free particle as
\begin{equation}\label{uuudewqas}
    (i\hbar~^t\mathfrak{D}_{F}^{\alpha})\psi_{F}^{\alpha}(t,\textbf{w}(v))=~\chi_{F}(t)\frac{- \hbar^2}{2 m}(~^{w_{i}}\mathfrak{D}_{F}^{\alpha})^2\psi_{F}^{\alpha}(t,\textbf{w}(v))).
\end{equation}
The Eq. (\ref{uuudewqas}) leads to the definition of the Hamiltonian and momentum operator on fractal curves as
\begin{equation}\label{s}
    \hat{H}_{F}^{\alpha}=i\hbar~^t\mathfrak{D}_{F}^{\alpha}~~~\hat{P}_{F}^{\alpha}=-i\hbar\mathfrak{\nabla}_{F}^{\alpha}
\end{equation}
The solution of Eq. (\ref{uuudewqas})  can be find using conjugate equation as
\begin{equation}\label{dewqas}
    i\hbar~\frac{\partial \theta(t,\xi)}{\partial t}=\frac{- \hbar^2}{2 m}\frac{ \partial^2}{\partial \xi^2}\theta(t,\xi)~~~\theta(\xi,t)=\phi[\psi_{F}^{\alpha}(t,\textbf{w}(v)))].
\end{equation}
Since the solution Eq. (\ref{dewqas}) is
\begin{equation}\label{derswq}
    \theta(t,\xi)=(A e^{i k\xi}+B e^{-i k\xi})e^{-i \beta t},
\end{equation}
where $k=\frac{\sqrt{2mE}}{\hbar}$ and $\omega=\frac{E}{\hbar}$ are constants. Now by applying $\phi^{-1}$ we have the solutions as
\begin{equation}\label{dertgbderswq}
    \psi_{F}^{\alpha}(t,\textbf{w}(v)))=(A e^{i k S_{F}^{\alpha}(v) }+B e^{-i k S_{F}^{\alpha}(v)})e^{-i \beta S_{F}^{\alpha}(t)}.
\end{equation}
 It is straight forward to extended to the case of a free particle  to the motion involving the potential. In this case the Lagrangian will be $L_{F}^{\alpha}=T_{F}^{\alpha}-V_{F}^{\alpha}(t, \textbf{w}(v))$. By substituting the Lagrangian in the Eq. (\ref{rewq}) one can derive the Schr\"{o}dinger equation as
\begin{align}\label{xxxx}
      \psi_{F}^{\alpha}(t,\textbf{w}(v))+\epsilon (~^t\mathfrak{D}_{F}^{\alpha})\psi_{F}^{\alpha}(t,\textbf{w}(v))\simeq \chi_{F}(t)\int_{-\infty}^{+\infty}\frac{1}{A}\exp[\frac{i}{\hbar}\frac{m \delta^2}{2\epsilon}]
      [1-\frac{i \epsilon}{\hbar}V_{F}^{\alpha}(t, \textbf{w}(v))](
    \psi_{F}^{\alpha}(t,\textbf{w}(v))\nonumber \\+ \delta (~^{w_{i}}\mathfrak{D}_{F}^{\alpha})\psi_{F}^{\alpha}(t,\textbf{w}(v)))+\frac{\delta^2}{2}
    (~^{w_{i}}\mathfrak{D}_{F}^{\alpha})^2\psi_{F}^{\alpha}(t,\textbf{w}(v))) d_{F}^{\alpha} \delta.
\end{align}
The same manner we worked out above  the Eq. (\ref{xxxx}) becomes
\begin{equation}\label{schrode}
    (i\hbar~^t\mathfrak{D}_{F}^{\alpha})\psi_{F}^{\alpha}(t,\textbf{w}(v))=~\chi_{F}(t)\frac{- \hbar^2}{2 m}(~^{w_{i}}\mathfrak{D}_{F}^{\alpha})^2\psi_{F}^{\alpha}(t,\textbf{w}(v)))+ \chi_{F}(t)V_{F}^{\alpha}(t, \textbf{w}(v))\psi_{F}^{\alpha}(t,\textbf{w}(v))
\end{equation}
\section{ Continuity equation and probability on fractal \label{4-s}}
It is well known that the continuity equation is a important concept in quantum mechanics. Therefor, the probability density on the fractal for a particle is defined as
\begin{equation}\label{dshro}
    \rho_{F}^{\alpha}(t, \textbf{w}(v))=(~^{*}\psi_{F}^{\alpha}(t,\textbf{w}(v)))~\psi_{F}^{\alpha}(t,\textbf{w}(v)).
\end{equation}
The complex conjugate wave function of Eq. (\ref{schrode}) is
\begin{equation}\label{schrofd}
    (-i\hbar~^t\mathfrak{D}_{F}^{\alpha})~^{*}\psi_{F}^{\alpha}(t,\textbf{w}(v))=~\chi_{F}(t)\frac{- \hbar^2}{2 m}(~^{w_{i}}\mathfrak{D}_{F}^{\alpha})^2~^{*}\psi_{F}^{\alpha}(t,\textbf{w}(v)))+ \chi_{F}(t)V(t, \textbf{w}(v))~^{*}\psi_{F}^{\alpha}(t,\textbf{w}(v))
\end{equation}
where $V_{F}^{\alpha}(t, \textbf{w}(v))=~^{*}V_{F}^{\alpha}(t, \textbf{w}(v))$. Applying this identity is given below
\begin{equation}\label{dsexz}
    ^t\mathfrak{D}_{F}^{\alpha}(\psi_{F}^{\alpha}(t,\textbf{w}(v))~^{*}\psi_{F}^{\alpha}(t,\textbf{w}(v)))=~
    ^t\mathfrak{D}_{F}^{\alpha}(\psi_{F}^{\alpha}(t,\textbf{w}(v)))^{*}\psi_{F}^{\alpha}(t,\textbf{w}(v))+
    \psi_{F}^{\alpha}(t,\textbf{w}(v))~^t\mathfrak{D}_{F}^{\alpha}(~^{*}\psi_{F}^{\alpha}(t,\textbf{w}(v))),
\end{equation}
and substituting  Eq. (\ref{schrode})  and  Eq. (\ref{schrofd}),  into Eq.(\ref{dsexz}) yield us
\begin{align}\label{qqqqq}
    i \hbar~^t\mathfrak{D}_{F}^{\alpha}\rho_{F}^{\alpha}(t, \textbf{w}(v))=\chi_{F}(t) \frac{\hbar^2}{2m}[\psi_{F}^{\alpha}(t,\textbf{w}(v))(~^{w_{i}}\mathfrak{D}_{F}^{\alpha})^2~^{*}\psi_{F}^{\alpha}(t,\textbf{w}(v))\nonumber\\
    -~^{*}\psi_{F}^{\alpha}(t,\textbf{w}(v))(~^{w_{i}}\mathfrak{D}_{F}^{\alpha})^2~\psi_{F}^{\alpha}(t,\textbf{w}(v))].
\end{align}
As a consequence the definition of  a probability current density on fractal curve is
\begin{align}\label{sa}
    J_{F}^{\alpha}(t,\textbf{w}(v))=\chi_{F}(t) \frac{\hbar}{2 m i}[\psi_{F}^{\alpha}(t,\textbf{w}(v))(~^{w_{i}}\mathfrak{D}_{F}^{\alpha})^2~^{*}\psi_{F}^{\alpha}(t,\textbf{w}(v))\nonumber\\
    -~^{*}\psi_{F}^{\alpha}(t,\textbf{w}(v))(~^{w_{i}}\mathfrak{D}_{F}^{\alpha})^2~\psi_{F}^{\alpha}(t,\textbf{w}(v))]
\end{align}
In the following table correspondence between standard quantum mechanics and generalized quantum framework  is presented.\\

\begin{tabular}{llr}
\hline
\multicolumn{2}{c}{Comparison between Standard Quantum and Quantum on Fractals} \\
\cline{1-2}
Postulates & ~~~~~~~~~~~~~~~~~~~Standard Quantum   & Quantum on Fractals \\
\hline
State       &~~~~~~~~~~~~~~~~~~~~~~~~~~~~~~$\psi(t,x)$   &  $\psi_{F}^{\alpha}(t,\textbf{w}(v))$     \\
Hamiltonian &~~~~~~~~~~~~~~~~~~~~~~~~~~~~~~~$i\hbar\frac{\partial}{\partial x}$       & $i\hbar~^t\mathfrak{D}_{F}^{\alpha}$~    \\
Momentum    &~~~~~~~~~~~~~~~~~~~~~~~~~~~~~~~~$ -i\hbar\nabla$  & $-i\hbar\mathfrak{\nabla}_{F}^{\alpha}$      \\
\hline
\end{tabular}

\section{Conclusion\label{5-s}}
The calculus on sets, vector space and manifold is used in the classical, quantum mechanics and general relativity respectively.~The geometry has important role in this generalization and modeling the physical phenomena. Recently, fractal geometry has been suggested by Mandelbrot.~So the calculus on them has been suggest by many researcher but it is still an open problem.~In this work we have studied the calculus on fractal curves.~Since the path integral in  Feynman formulation is fractal so that is motivated us to suggest this generalization.~This framework can suggest correct way for  obtaining Schr\"{o}dinger equation from Fyenman path quantum mechanics.\\

\section*{Acknowledgements}
 One of the authors (AKG)  would like to thank Professor A. D. Gangal for useful discussion on this topic  during the period of time he was in Pune University.

%
%

\end{document}